\documentclass[graybox]{svmult}

\usepackage{mathptmx}       % selects Times Roman as basic font
\usepackage{helvet}         % selects Helvetica as sans-serif font
\usepackage{courier}        % selects Courier as typewriter font
\usepackage{type1cm}        % activate if the above 3 fonts are
\usepackage{makeidx}         % allows index generation
\usepackage{graphicx}        % standard LaTeX graphics tool
\usepackage{multicol}        % used for the two-column index
\usepackage[bottom]{footmisc}% places footnotes at page bottom
\usepackage{comment}
\usepackage{caption}
\usepackage{subcaption}
\usepackage{apacite}
\makeindex             % used for the subject index
\begin{document}

\title*{Exogenous Shocks Lead to Increased Responsiveness and Shifts in Sentimental Resilience in Online Discussions}
\titlerunning{Exogenous Shocks Lead to Shifts in Online Responsiveness and Sentimental Resilience}

\author{Chathika Gunaratne, Subash K. Ray, Caroline Louren\c{c}o Alves, and Maria Waldl}
%\authorrunning{Gunarate, Ray, Alves, \& Waldl}
\institute{Chathika Gunaratne \at University of Central Florida, Orlando, FL, USA \email{chathika.gunaratne@ucf.edu}
\and Subash K. Ray \at New Jersey Institute of Technology \\\& Rutgers University - Newark, Newark, NJ, USA \email{sr523@njit.edu}
\and Caroline Louren\c{c}o Alves \at University of Sao Paulo, Sao Paulo, Brazil. \email{caroline.lourenco.alves@usp.br}
\and Maria Waldl \at University of Vienna, Vienna, Austrai \email{maria@tbi.univie.ac.at}}
\maketitle

\abstract*{The effects of real-world events on the dynamics and sentiment expressed through online conversations is not entirely understood. In particular, the dynamics of highly polarized communities, deeply invested in the outcome of a particular event naturally tend to have a contrasting emotional sensitivity to exogenous events.
\newline\indent 
In this study, we analyze Twitter conversations during the Mexico vs Germany group match of the 2018 FIFA World Cup, and investigate the effect of the live game outcomes on the conversation dynamics and sentiment.
%We analyze the influence of the sentiment expressed through the game commentary on the volume, virality, responsiveness, and sentiment expressed in the online discussions over time. 
We find that the exogenous events influence conversation volume and virality less, while having high influence on user responsiveness.
Interestingly, we observe a shift in the influence that exogenous events have on fans immediately following the only goal scored in the game. The emotional resilience of fans of the advantaged team increased following this exogenous shock. In contrast, the sentiment of the disadvantaged team was left more susceptible to further exogenous events following the exogenous shock. 
These results support the fact that user engagement and emotional resilience of an online population holding highly polarized stances can be manipulated through an exogenous event of high importance.}

\abstract{The effects of real-world events on the dynamics and sentiment expressed through online conversations is not entirely understood. In particular, the dynamics of highly polarized communities, deeply invested in the outcome of a particular event naturally tend to have a contrasting emotional sensitivity to exogenous events.
\newline\indent 
In this study, we analyze Twitter conversations during the Mexico vs Germany group match of the 2018 FIFA World Cup, and investigate the effect of the live game outcomes on the conversation dynamics and sentiment.
%We analyze the influence of the sentiment expressed through the game commentary on the volume, virality, responsiveness, and sentiment expressed in the online discussions over time. 
We find that the exogenous events influence conversation volume and virality less, while having high influence on user responsiveness.
Interestingly, we observe a shift in the influence that exogenous events have on fans immediately following the only goal scored in the game. The emotional resilience of fans of the advantaged team increased following this exogenous shock. In contrast, the sentiment of the disadvantaged team was left more susceptible to further exogenous events following the exogenous shock. 
These results support the fact that user engagement and emotional resilience of an online population holding highly polarized stances can be manipulated through an exogenous event of high importance.}

\section{Introduction}
\label{section:introduction}
Online social networks offer a medium for the discussion of real world events at global scales, transcending the challenges experienced by other forms of human communication. Individuals with completely different experiences and cultural backgrounds and opposing opinions on a topic may now be spontaneously exposed to one another when discussing a common topic of interest. This results in complex conversation dynamics arising between the social media users due to the many factors driving the expression of ideas and/or opinions, compounded by the constant intrusion of opposing information or opinions. 

Online social media discussions are typically initiated by an event exogenous to the social media platform. If a sufficient number and rate of responses to an initial post is received, then the original post (or root) is able to accumulate responses over time, endogenously growing into and online conversation, also referred to as a information cascade. Often, new information is generated in response to an exogenous event in the form of replies to a tweet about the event. This influx of new information, contributes in the continuation of a conversation, and expression of opinions and positions.%This influx of new information, along with the replies expressing opinions and positions, contribute in the continuation of the conversation. 

Understanding the extent to which exogenous events affect online conversation dynamics in comparison to the endogenous forces of discussion is the focus of this study. We expect the impact of exogenous information on online conversation dynamics to be stronger when considering a rapidly discussed topic, with several unprecedented and high-impact events, being discussed among two groups of a highly-invested and highly-polarized populations, such as a live sports game or an election. Accordingly, we choose the  Mexico vs Germany group match of the 2018 FIFA World Cup as our topic of interest, which ended with an unprecedented win by Mexico scoring 1-0. We analyze the conversations that were exchanged over Twitter right before, during, and after the single goal scored over the course of the game. We selected this game in particular, due to the win for Mexico over the team with a stronger historical record, and for the lingual difference between fans of the two teams, for the ease of analysis. The Twitter discussions during the course of this game provide a unique opportunity to study a polarized population engaged in competitive discussion, in isolation, over a fixed time period, along with the explicit specification of exogenous shocks important to the community in the form of the game commentary transcript.

In this study, we analyze the effect that the exogenous events, in particular the game events reported in the game transcript, have on the online conversation dynamics and the sentiment expressed by fans of both teams. Conversation dynamics are quantified via three measurements: 1) \textit{volume}, the total number of replies or retweets to the root tweet or replies or retweets of the resulting conversation, 2) \textit{virality}, the Weiner index of the conversation tree resulting from the root tweet, and 3) \textit{responsiveness}, the rate at which replies and retweets are accumulated by the conversation resulting from the root tweet. These measures have been used in similar investigations in the literature \cite{choi2015characterizing}. Furthermore, we performed sentiment analysis \cite{kouloumpis2011twitter} on both the Twitter conversations and game commentary to obtain polarity scores, to quantify sentiment. Finally, we performed transfer entropy analysis on the resulting time-series data to investigate the influence that the game events had on the conversation dynamics and the influence that the commentary sentiment had on the sentiment expressed in the conversations.

Our results indicate that the popularity of the root user had greater influence than the game events on the volume and virality of the Twitter conversations. However, the game events were found to have stronger influence overall on the responsiveness to the conversations. Further, game events at around 7 seconds prior to the conversation had the highest influence on the conversation volume, while game events around 22 to 29 seconds had highest influence on the virality of conversations overall. Interestingly, game events by Germany had a higher influence on the responsiveness than the game events by Mexico, despite Mexico winning the match. The commentary sentiment is shown to have higher influence coming up to the goal than after the goal overall. Finally, we see that users tweeting in German language have higher sentimental influence after the goal than prior to the goal by Mexico, indicating that there was anticipation following the goal by Mexico, that Germany would score in the game. 

\section{Background}
Sentiment analysis is a growing area of Natural Language Processing (NLP) and refers to the extraction and classification of a text primarily available on the web, like social media posts, news articles and blog posts, to determine the opinions, emotions and attitude of people \cite{wilson2005recognizing}\cite{liu2012sentiment}. The purpose of this type of analysis is to subjectify people's sentiments about a news item (for e.g., a political situation, sports and weather), reviews on books, movies or consumer product, which are carried out by political analysts, marketeers and companies. A typical approach to sentiment analysis is to start with a lexicon of positive, negative and neutral words and phrases \cite{wilson2005recognizing}; thess entries are labeled with their polarity. In particular, Twitter has been the one of the most popular social networking website generating enormous amount of data because of its millions of followers that are increasing by the day \cite{sehgal2018real}. Therefore, sentiment analysis on  data has been an effective tool in discerning public attitude towards a certain topic \cite{saif2012semantic}. 

Choi et al \cite{choi2015characterizing} use trees to represent the structure of conversations. Vertices of these trees represent comments by users, linked by edges which represent a response to another comment. The author's then characterize these trees with measures of volume, virality, and responsiveness.
%To characterize these trees the following measures can be used: ---- Chathika: Addressed% 

In addition, in order to compute the information transfer from the sentiments of the events happening during the game to the sentiments of the tweets posted by people on, we used transfer entropy. Transfer entropy is a model-free, information-theoretic tool that measures the amount of directed predictive information flow between two time-series when one of the time-series is hypothesized to influence the other \cite{PhysRevLett.85.461}. In specific, transfer entropy from a time-series X to a time-series Y (or $T_{X\rightarrow Y}$) is the measure of reduction in uncertainty (or increase in predictability) in the future values of Y by using the knowledge of the past values of X given the past values of Y.  Transfer entropy $T_{X\rightarrow Y}$ can be calculated using the equation \ref{eqution:transfer}.

\begin{equation}
\label{eqution:transfer}
T_{X\rightarrow Y} = \sum_{x_{i+1},x_{i}^{(k)},y_{i}}p(x_{i+1},x_{i}^{(k)},y_{i}) log_{2} \frac{p(x_{i+1},y_{i}|x_{i}^{(k)})}{p(x_{i+1}|x_{i}^{(k)})p(y_{i}|x_{i}^{(k)})}.
\end{equation}

Here, \textit{X} and \textit{Y} are time-series, with $x_{i}^{(k)} = {x_{(i-k+1)},x_{(i-k+2)},...,x_i}$  represents the \textit{k} successive past values of time-series \textit{X} at time \textit{i}  and including the value at the current time (i.e.,  $x_i$).

\section{Methodology}
\subsection{Data}
\begin{figure}[t]
\sidecaption
\includegraphics[width=\linewidth]{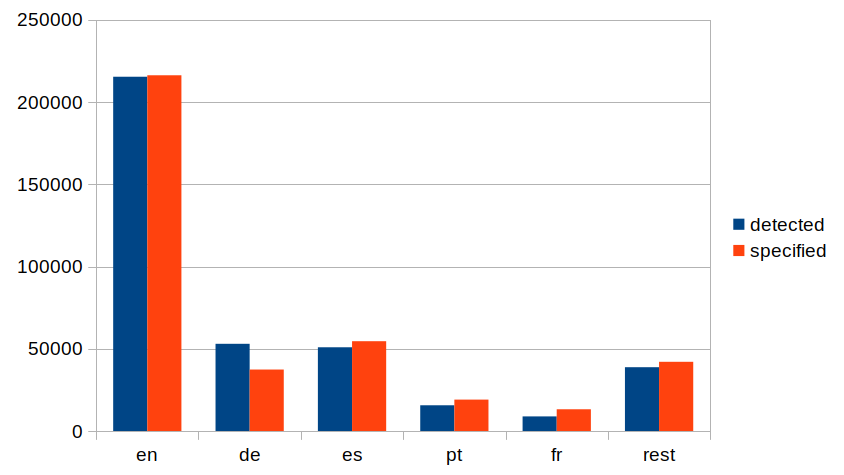}
\caption{Histogram of language of collected tweets as specified by Twitter API (red) and detected (blue) through python-langdetect library. English (en), German (de) and Spanish (es) are the most common, followed by Portuguese (pt) and French (fr).}
\label{fig:languse}
\end{figure}

We collected 382042 tweets by fans posting on the Mexico vs Germany opening game of FIFA 2018 using  Developer API. Tweets were collected 30 minutes before the game started to 30 minutes after the game ended, from 2018-06-17 14:30:00 to 2018-06-17 17:00:00. Tweets that mentioned the hashtags: \#GERMEX, \#FIFA, \#Germany, or \#Mexico, are were created during this period were collected. The collected tweets were in multiple languages, from which English, Spanish, and German were the most commonly used as shown in Fig. \ref{fig:languse}. The tweet texts were translated to English using Amazon's AWS Traslate API with AWS EC2. The tweet texts were then stemmed and run through the Vader sentiment analysis algorithm to extract polarity scores. In addition to collecting the tweets, the game transcript was obtained to represent the effect of the events of the game on the fans \footnote{http://www.espn.com/soccer/commentary?gameId=498193}.

\subsection{Conversation Dynamics}
	Conversations were treated as trees of response tweets to an originating root tweet. A conversation could have none to many replies, be shallow or deep/viral, and could call low to high response rates, respectively. These conversation dynamics were measured by three properties similar to \cite{choi2015characterizing}, volume, virality, and responsiveness. Volume was simply measured as the total number of tweet replies, retweets, or quoted tweets that a root tweet received.

\begin{figure}[tbh!] 
    \centering
    \begin{subfigure}{0.49\linewidth}
        \includegraphics[width=\linewidth]{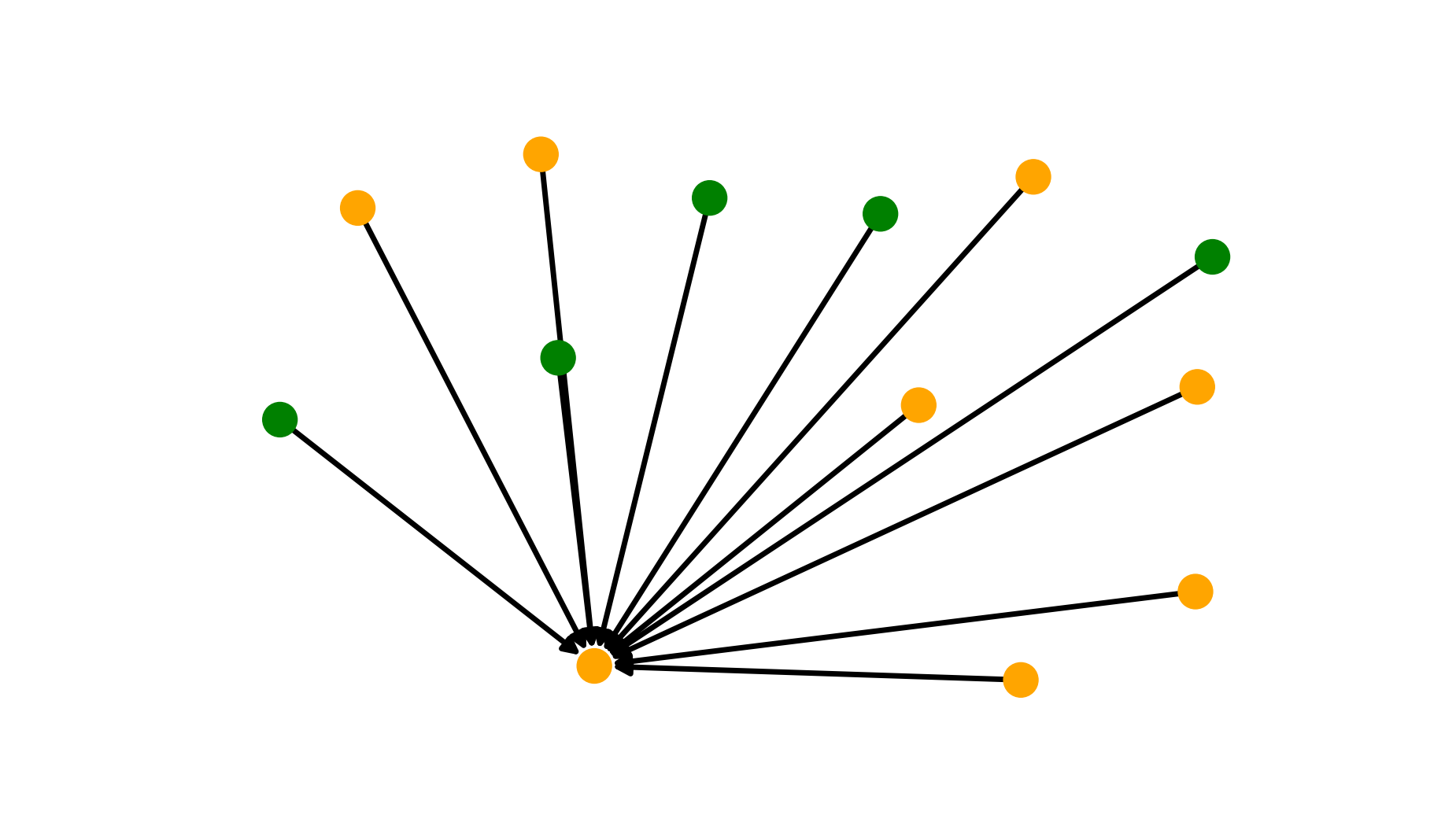}
        \caption{Large conversation with low virality.}
        \label{fig:viralitylow}
    \end{subfigure}
    \begin{subfigure}{0.49\linewidth}
        \includegraphics[width=\linewidth]{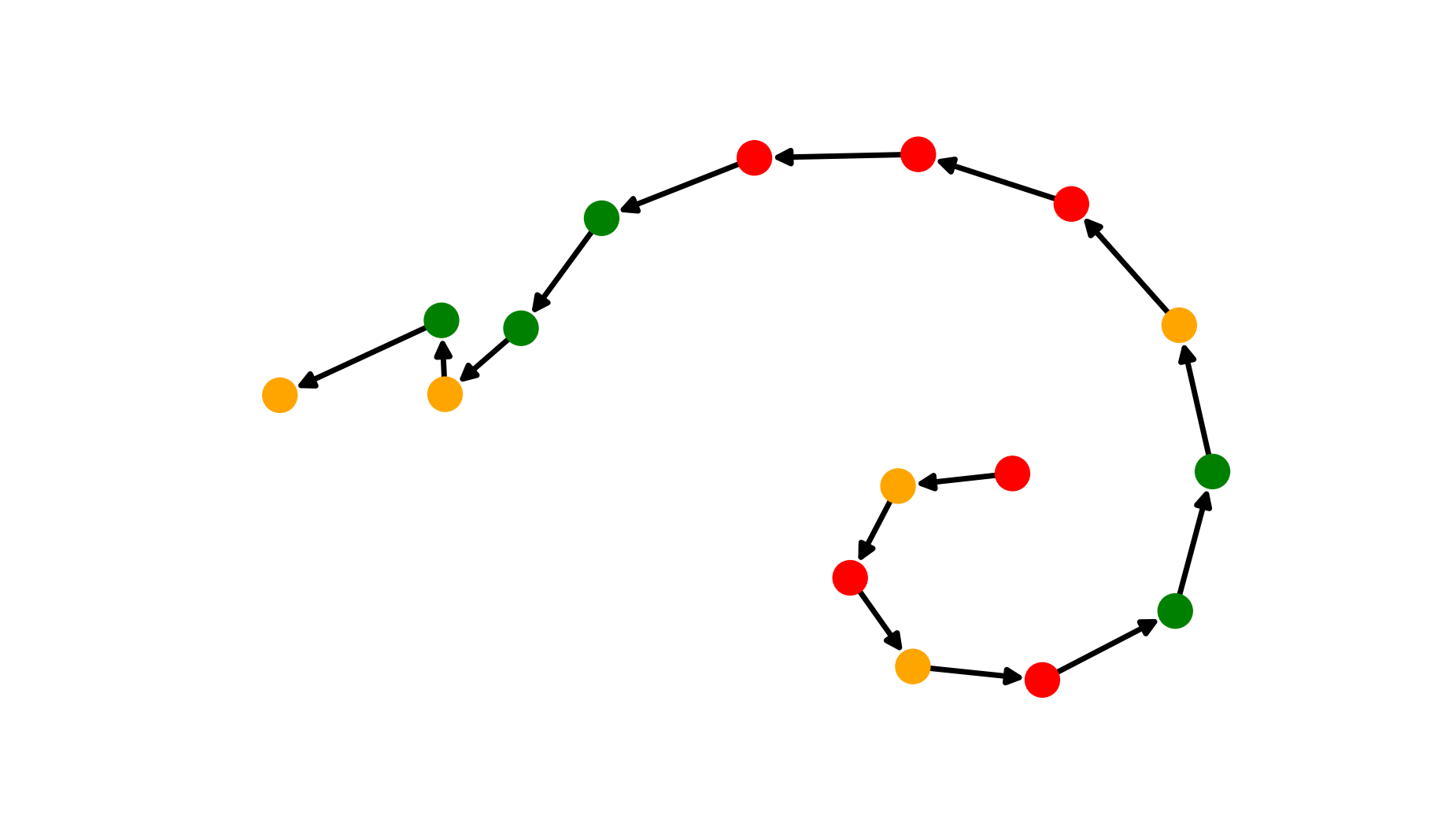}
        \caption{Large conversation with high virality.}
        \label{fig:viralityhigh}
    \end{subfigure}
\caption{Visual examples of two large conversations in the dataset, with \ref{fig:viralitylow} low virality and \ref{fig:viralityhigh} high virality, as determined through the Wiener index. Sentiment polarity scores of each tweet has been depicted as negative (red), neutral (yellow), or positive (green).}
\label{fig:viralityexample}
\end{figure}

Secondly, structural virality indicated the shape of the conversation by the depth of the conversation tree with respect to its width. A tree with high virality would have a long chain of replies (Fig \ref{fig:viralityexample}), compared to replies centered around a single tweet. As in \cite{choi2015characterizing}, structural virality was calculated via the Wiener Index \cite{wiener1947structural}, a graph property that can be defined as the average distance between all pairs of nodes in a diffusion tree \cite{goel2015structural}, \textit{v(T)}, described in equation \ref{equation:virality}.
\begin{equation}
\label{equation:virality}
v(T)= \frac{1}{n(n-1)}\sum_{i=1}^{n}\sum_{j=1}^{n} d_{ij},
\end{equation}
where $d_{ij}$ denotes the length of the shortest path between nodes \textit{i} and \textit{j}; and \textit{n} the number of all nodes.

Thirdly, Responsiveness indicated the rate at which replies would be posted to a parent tweet on a conversation. Responsiveness was calculated as the mean, inverse difference between the time of creation of the reply and the time of creation of the parent tweet, for all replies on a conversation. We used a resolution of seconds for the creation of timestamps, making the unit of responsiveness used in the study in terms of tweets per second.

\begin{figure}[tbh!]
\centering
\includegraphics[width=\linewidth]{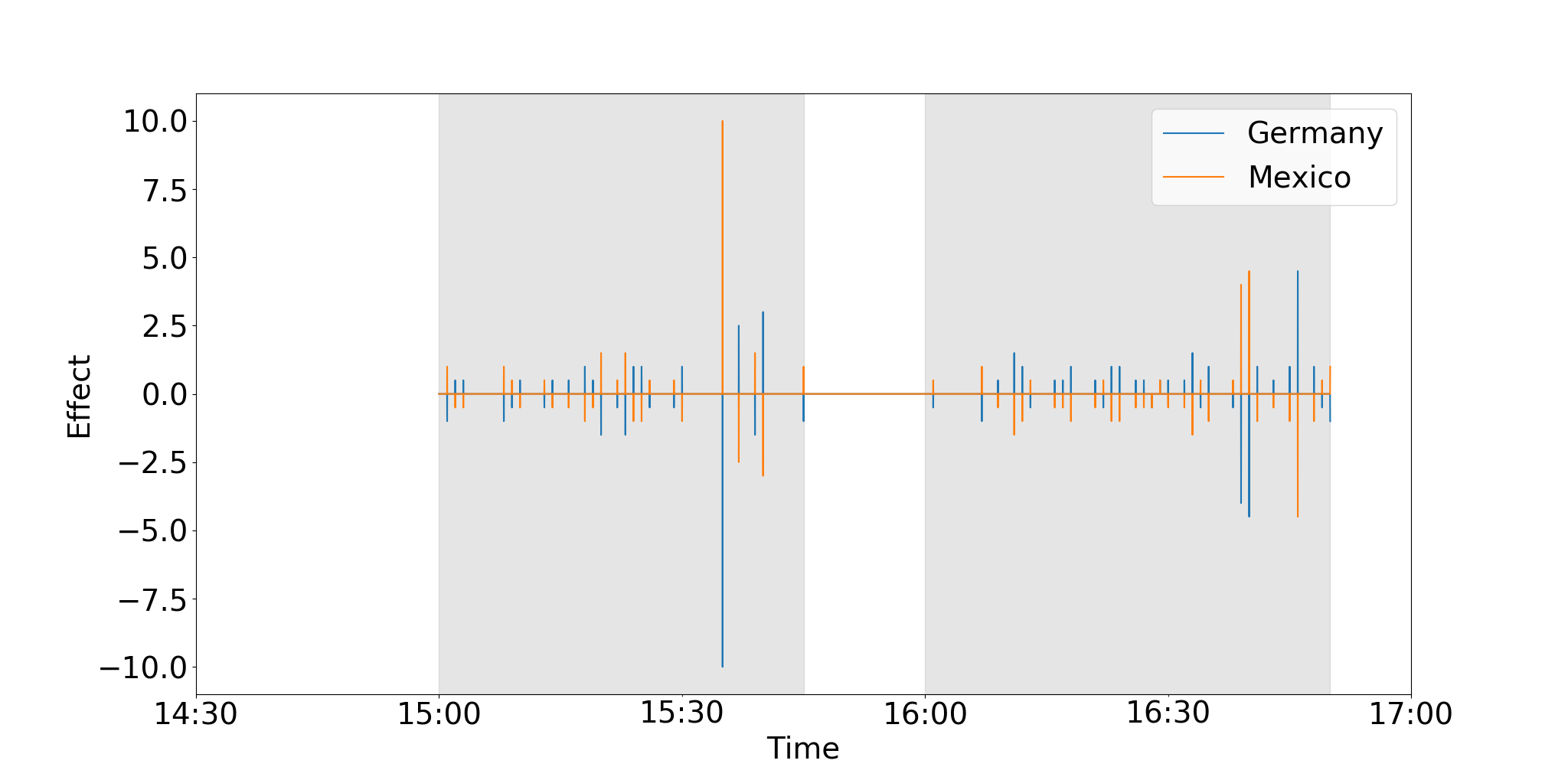}
\caption{The game events time-series generated by evaluation of the events reported in the game transcript. Both teams received scores for each event and the scores for Germany are denoted in blue, while those for Mexico are in orange.}
\label{fig:transcript}
\end{figure}

Finally, the transcript was used to generate a time-series of game event effects experienced by both the German and the Mexico fans, along with an overall total game events time-series. The game event effects time-series for both the German and Mexican teams were created by applying the following rules, to approximate the intensity of exogenous game events on either team due to the events reported on the transcript: a foul by a team results in -0.5 for that team and +0.5 for the receiving team, a saved goal results in +0.5 for the attacking team and -0.5 for the defending team, a goal results in +10 for the scoring team and -10 for the opposing team, finally, a yellow card results in -3 for the committing team and +3 for the opposing team. The resulting game events time-series is displayed in Fig. \ref{fig:transcript}.

\subsection{Sentiment Transfer}
\subsubsection{Data Pre-processing}
The sentiment analysis computation is performed using the Vader Sentiment analyzer for Python \footnote{https://github.com/cjhutto/vaderSentiment}\cite{gilbert2014vader}. The sentiment analysis provided scores indicating negative, neutral, positive sentiment of text, in addition to a combined compound sentiment score. We used the compound sentiment score for our analysis. The compound score provided by Vader is already normalized between -1 to 1.

We obtained the sentiment time-series by calculating the average sentiment at every minute of the game transcript and sentiments of the tweets ( see Figure \ref{fig:Sentiment time-series}. %We only considered tweets that were posted during the course of the game. 
We calculated the first derivative of both the time-series and obtained two new times series, i.e., $X = {x_1,...,x_n}$ and $Y = {y_1,...,y_n}$, corresponding to the game transcript and tweet sentiments, respectively. We assigned a value 3 (i.e.,  $x_i$  and $y_i=3$) when the derivative was positive, a value 2 (i.e., $x_i$  and $y_i=2$) when the derivative was equal to 0, and a value 1 (i.e., $x_i$  and $y_i=1$) when the derivative was negative. The two time-series were used as inputs for the measurement of transfer entropy explained in the section \ref{section:introduction}. 

\subsection{Transfer Entropy}
As already mentioned in section \ref{section:introduction}, we used transfer entropy to compute the information transfer from the sentiments of the events happening during the game to the sentiments of the tweets posted by people on .

Using this definition of transfer entropy, we defined a metric named Total Transfer Entropy (or TTE) to calculate the net information transfer between the time-series from the game transcript and tweets, which is given by equation \ref{eq:transfer}.
\begin{equation}
\label{eq:transfer}
TTE= T_{X\rightarrow Y} - T_{Y\rightarrow X}
\end{equation}

Where X is the sentiment time-series of the game transcript and Y is the sentiment time-series of the tweets. Therefore, TTE gives the aggregate information flow from the sentiments of the events happening during the game to the sentiments of the tweets posted by people on. This method has been used previously to infer leader-follower relationships in Zebrafish schools and direction of information transfer in Physarum polycephalum's membrane during food choice \cite{butail2016model,ray2019information}

We used TTE to measure the aggregate information flow from the game transcript to the sentiments for the duration of the game; and as well as during durations before and after the goal was scored (i.e., TTE  between the sentiments before and after the 33rd minute of the game time). Additionally, TTE  was measured from the game transcript to the sentiments of all tweets posted by people about the game on, and of people posting in languages English, Spanish and German.

\section{Results}

\subsection{Conversation Dynamics}
\begin{figure}
\centering
\includegraphics[width=\linewidth]{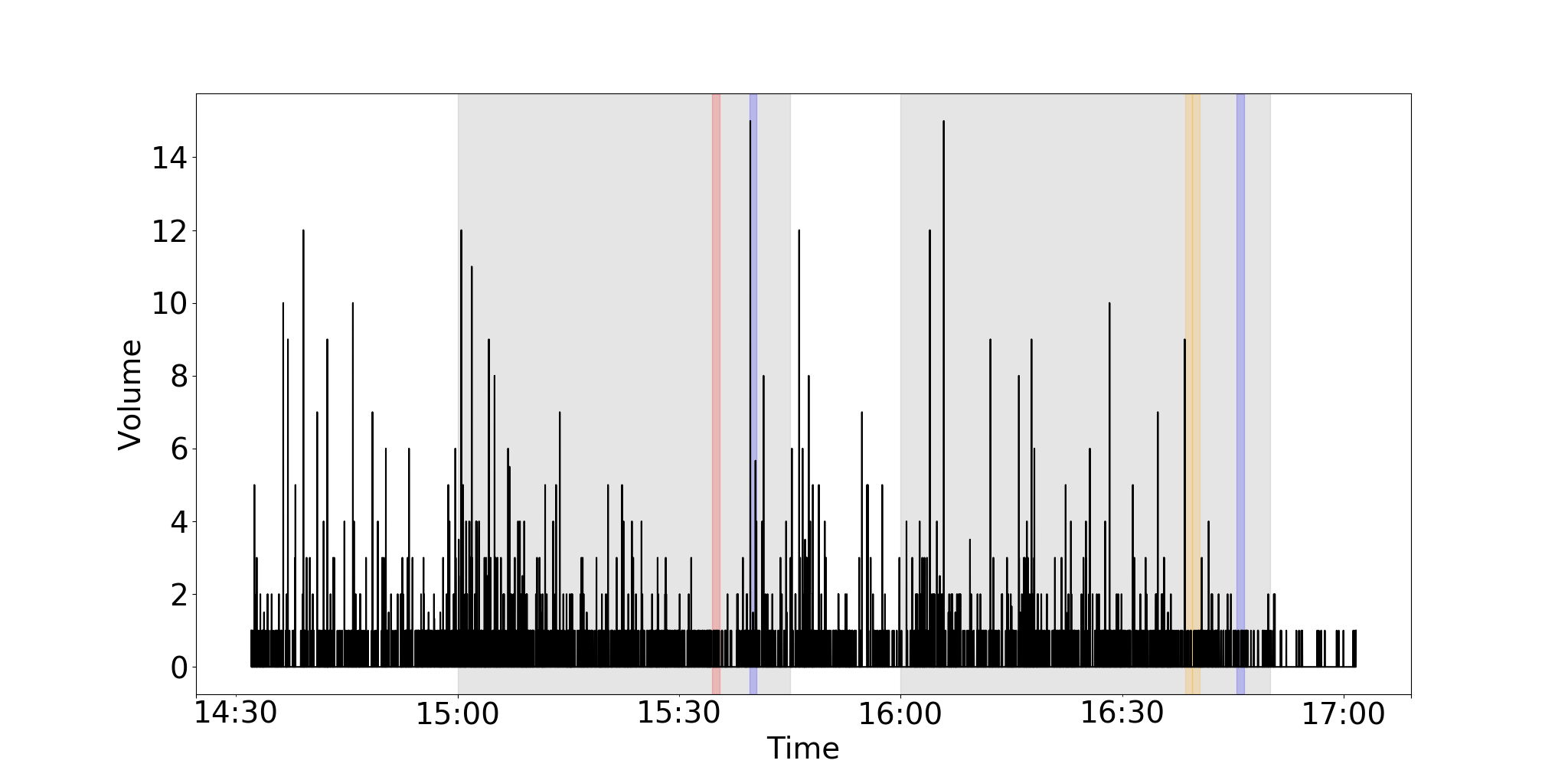}
\caption{The eventual volume of conversations originated by tweets over time. The grey areas depict the fist and second half of the game. The red bar indicates when Mexico scored the only goal of the game, while the purple and yellow bars indicate when Germany and Mexico players received yellow cards, respectively.}
\label{fig:volume}
\end{figure}

\begin{figure}
\centering
\includegraphics[width=\linewidth]{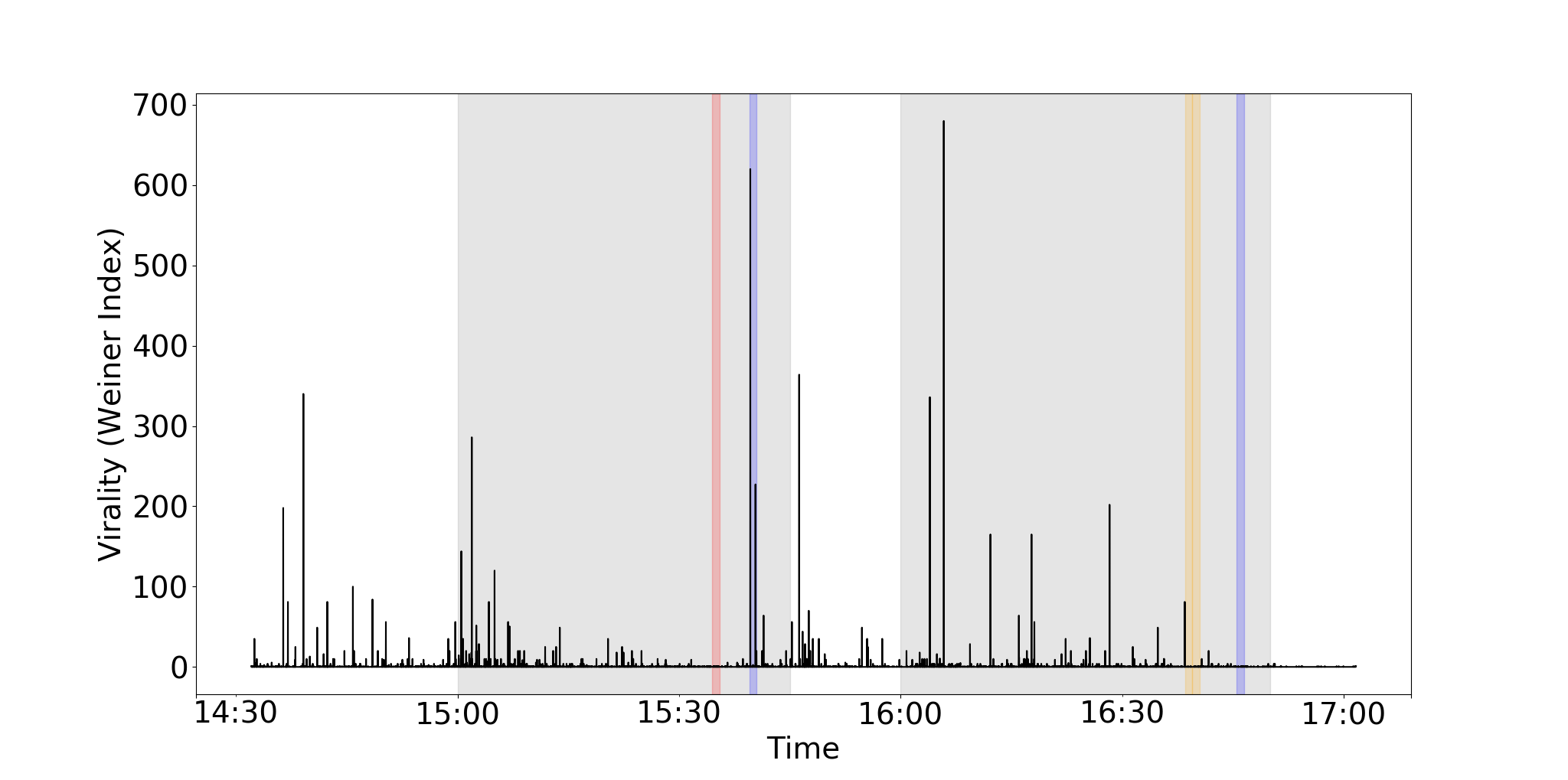}
\caption{The eventual virality (Weiner Index) of conversations originated by tweets over time. The grey areas depict the fist and second half of the game. The red bar indicates when Mexico scored the only goal of the game, while the purple and yellow bars indicate when Germany and Mexico players received yellow cards, respectively.}
\label{fig:virality}
\end{figure}

\begin{figure}
\centering
\includegraphics[width=\linewidth]{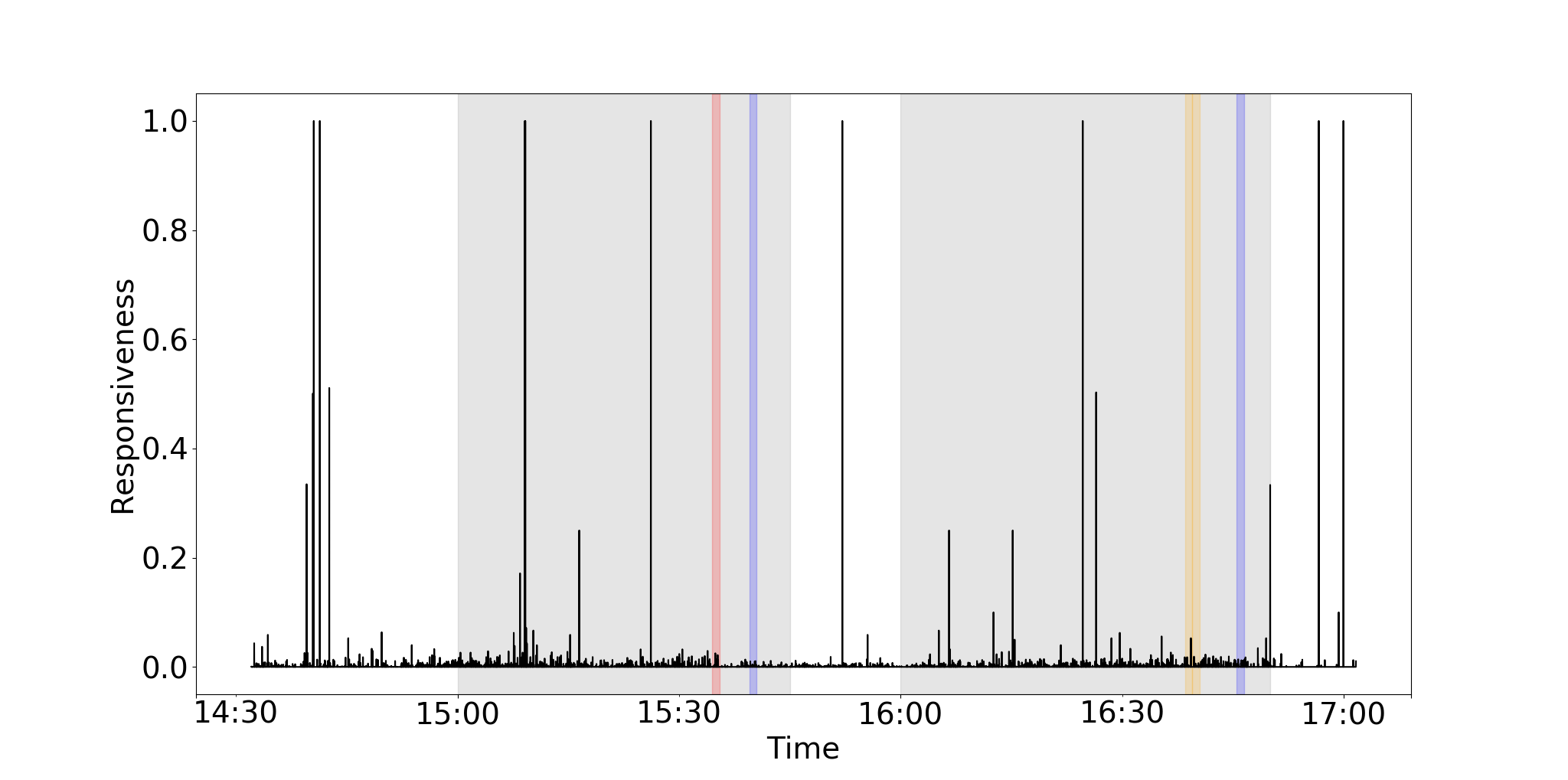}
\caption{The eventual responsiveness of conversations originated by tweets over time. The grey areas depict the fist and second half of the game. The red bar indicates when Mexico scored the only goal of the game, while the purple and yellow bars indicate when Germany and Mexico players received yellow cards, respectively.}
\label{fig:responsiveness}
\end{figure}

Shown in Figs. \ref{fig:volume}, \ref{fig:virality}, and \ref{fig:responsiveness}, are the time-series of volume, virality, and responsiveness of root tweets over the analyzed time period. Shaded in grey are the first and second halves of the game. Additionally, the time of occurrence of the goal (red), two yellow cards to Germany (purple) and two consecutive yellow cards to Mexico (yellow) have been marked. As shown in Figure 5, we observed that viral conversations initiated at the start of the first and second halves, and the end of the first half. %Additionally, it can be seen that one of the highest volume conversations and several viral conversations were initiated with the first yello 
In contrast, no such pattern to the game events is seen with responsiveness.	

\begin{figure}
\centering
\includegraphics[width= \linewidth]{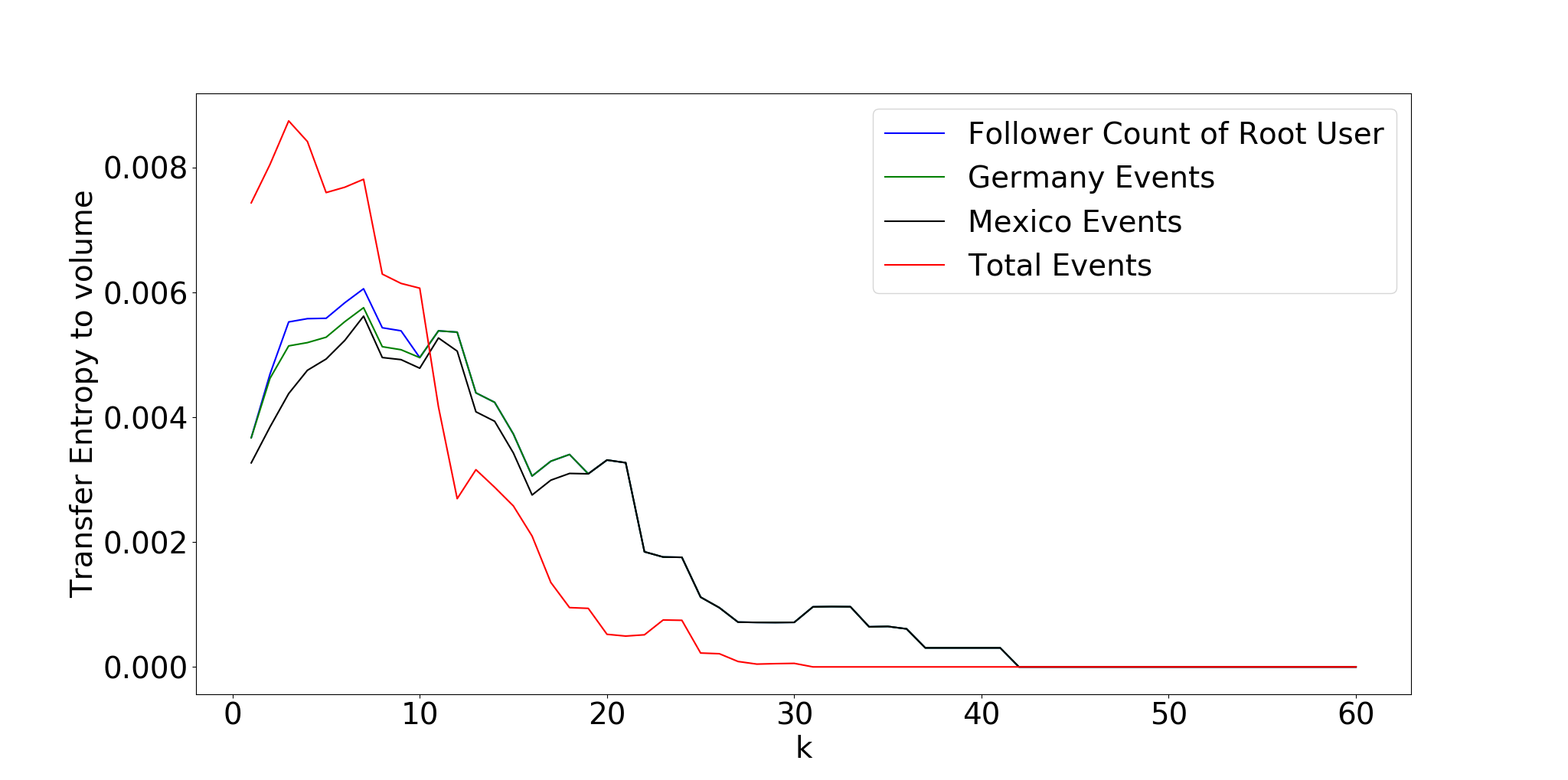}
\caption{Comparison of transfer entropy from external effects (Germany (blue), Mexico (green), and combined (black) game events) to the conversation volume time-series against the transfer entropy from the internal effect (the time-series of the number of followers of the conversation root user) over varying k.}
\label{fig:tevolume}
\end{figure}

\begin{figure}
\centering
\includegraphics[width= \linewidth]{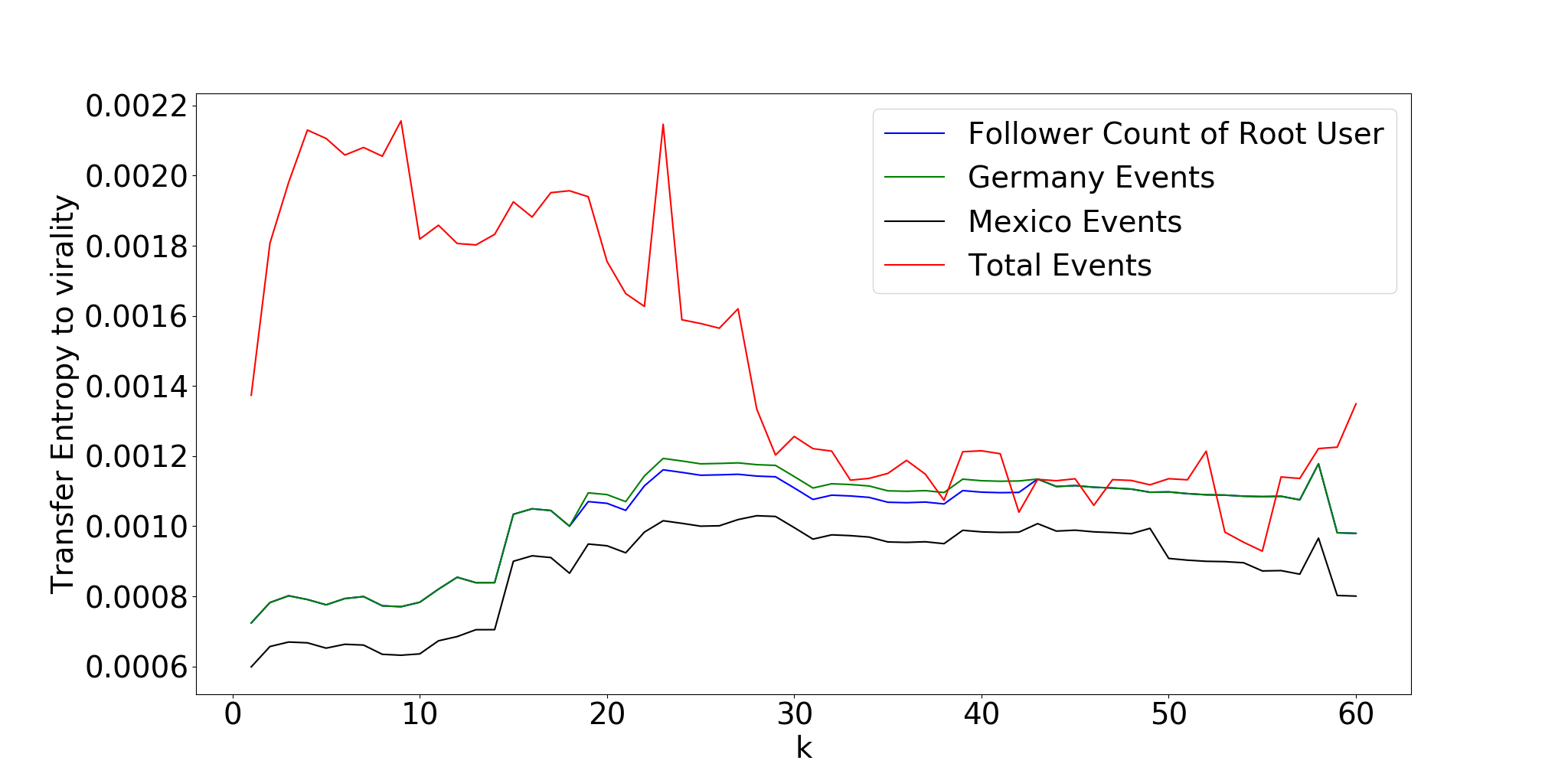}
\caption{Comparison of transfer entropy from external effects (Germany (blue), Mexico (green), and combined (black) game events) to the conversation virality time-series against the transfer entropy from the internal effect (the time-series of the number of followers of the conversation root user) over varying k.}
\label{fig:tevirality}
\end{figure}

\begin{figure}
\centering
\includegraphics[width= \linewidth]{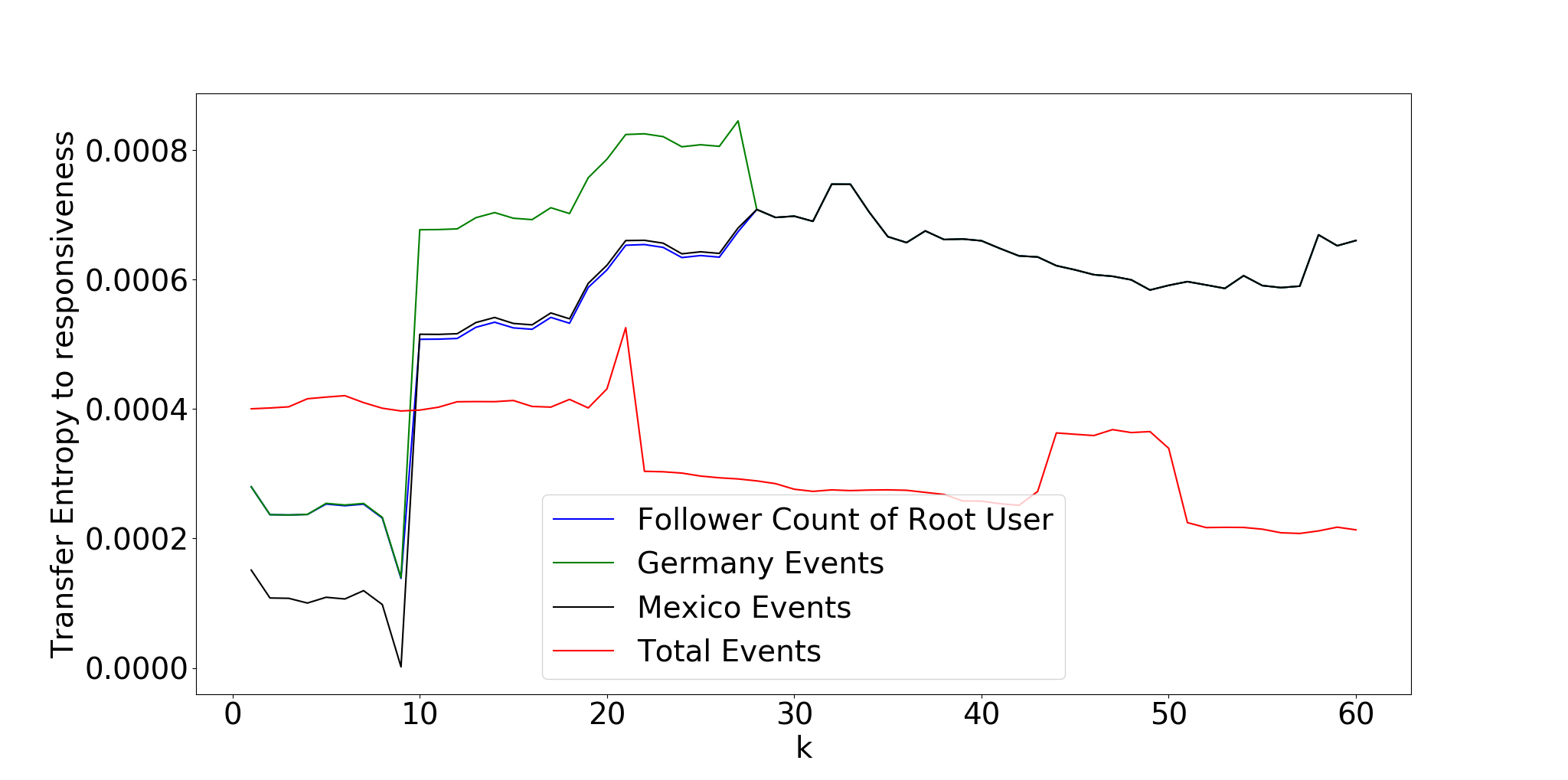}
\caption{Comparison of transfer entropy from external effects (Germany (blue), Mexico (green), and combined (black) game events) to the conversation responsiveness time-series against the transfer entropy from the internal effect (the time-series of the number of followers of the conversation root user) over varying k.}
\label{fig:teresponsiveness}
\end{figure}

Figs. \ref{fig:tevolume}, \ref{fig:tevirality}, and \ref{fig:teresponsiveness} plot the change in transfer entropy with the history length parameter, $k$, for the volume, virality, and responsiveness of root tweets, respectively, due to the performance of the two teams (Germany in blue, Mexico in green, and total in black). Additionally, the transfer entropy change with $k$ for the three measures by the endogenous factor, number of followers of the user posting the root tweet, is also plotted (red) for reference. It can be seen that there is a generally higher transfer entropy from the follower count of the root creators, than the team performance for volume and virality. However, when considering responsiveness the transfer entropy results from the game events are generally higher than from the follower count series of the conversation originators. 

Finally, we consider the history lengths at which the maximum transfer entropy values from the three game event time-series to the three conversation dynamics time-series were reported. The transfer entropy from all three game event series to conversation volume were maxmimal at a history length of $k=7s$; i.e.  the best estimate of information transfer from the game events to the conversation volume occurred in the past 7 seconds. The transfer entropy from all three game event series to conversation virality was maximal at a history length of approximately $22s<k<29s$, i.e. the best estimate of information transfer from the game events to the conversation virality occurred within the past 22 to 29 seconds. However, the maximal transfer entropy from the game events of the two teams to conversation responsiveness were different. Transfer entropy from the Mexican team events to conversation responsiveness was maximal at $21<k<27$, while the transfer entropy from the German team events to conversation responsiveness was maximal at $31<k<33$; i.e. the best estimate of information transfer from the game events by the Mexican team to the conversation dynamics occurred in the past 21 to 27 seconds, and that by German team occurred in the past 31 to 33 seconds.

%\begin{figure}
%\centering
%\includegraphics[width= \linewidth]{Figures/ViralityVsVolume.png}
%\end{figure}
%\begin{figure}
%\centering
%\includegraphics[width= \linewidth]{Figures/ViralityVsResponsiveness.png}
%\end{figure}

\subsection{Sentiment Transfer}

\begin{figure}
\centering
\includegraphics[width=\linewidth]
{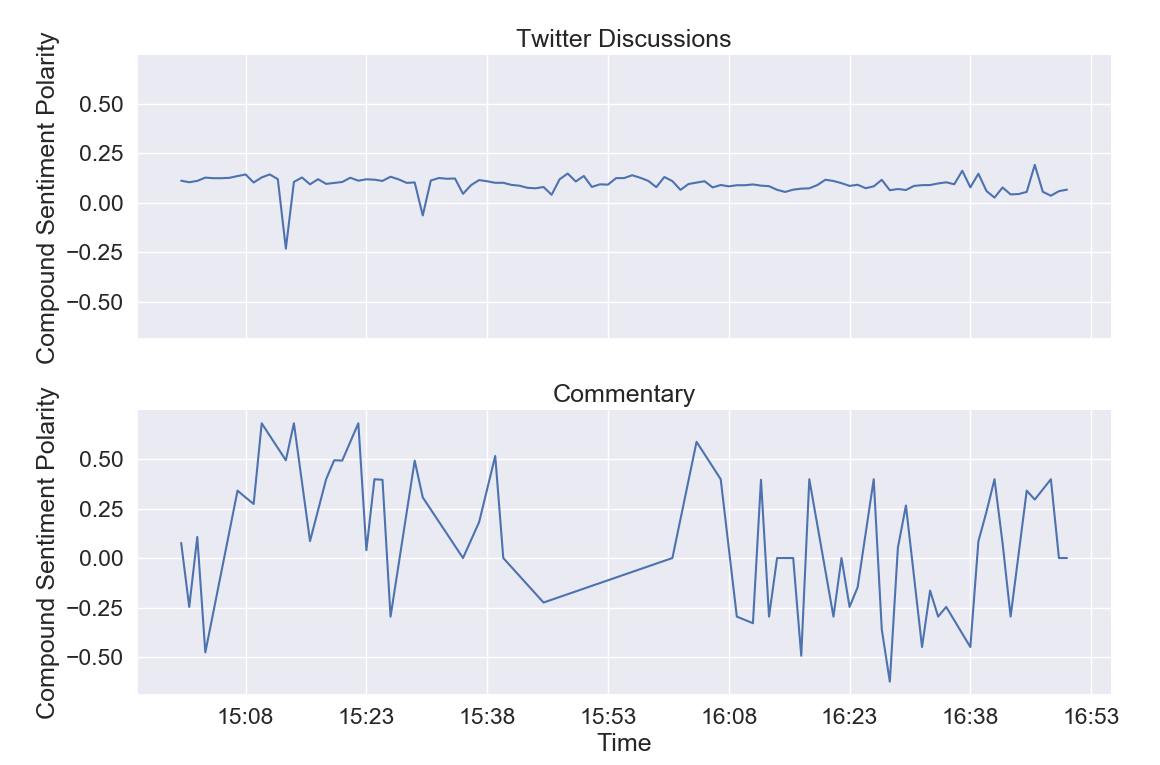}
\caption{Sentiment time-series of (A) Transcript and (B) Tweets posted when the game was in progress.}
\label{fig:Sentiment time-series}
\end{figure}

The sentiment time-series of the game transcript and tweets are shown in Fig. \ref{fig:Sentiment time-series}. The sentiment of the tweets generally stayed in the positive region whereas the sentiments were equally spread across positive and negative regions of the transcript sentiment time-series of the transcript. 

\subsection{Total Transfer Entropy}
\begin{figure}[tbh!]
\centering
\includegraphics[width=\linewidth]
{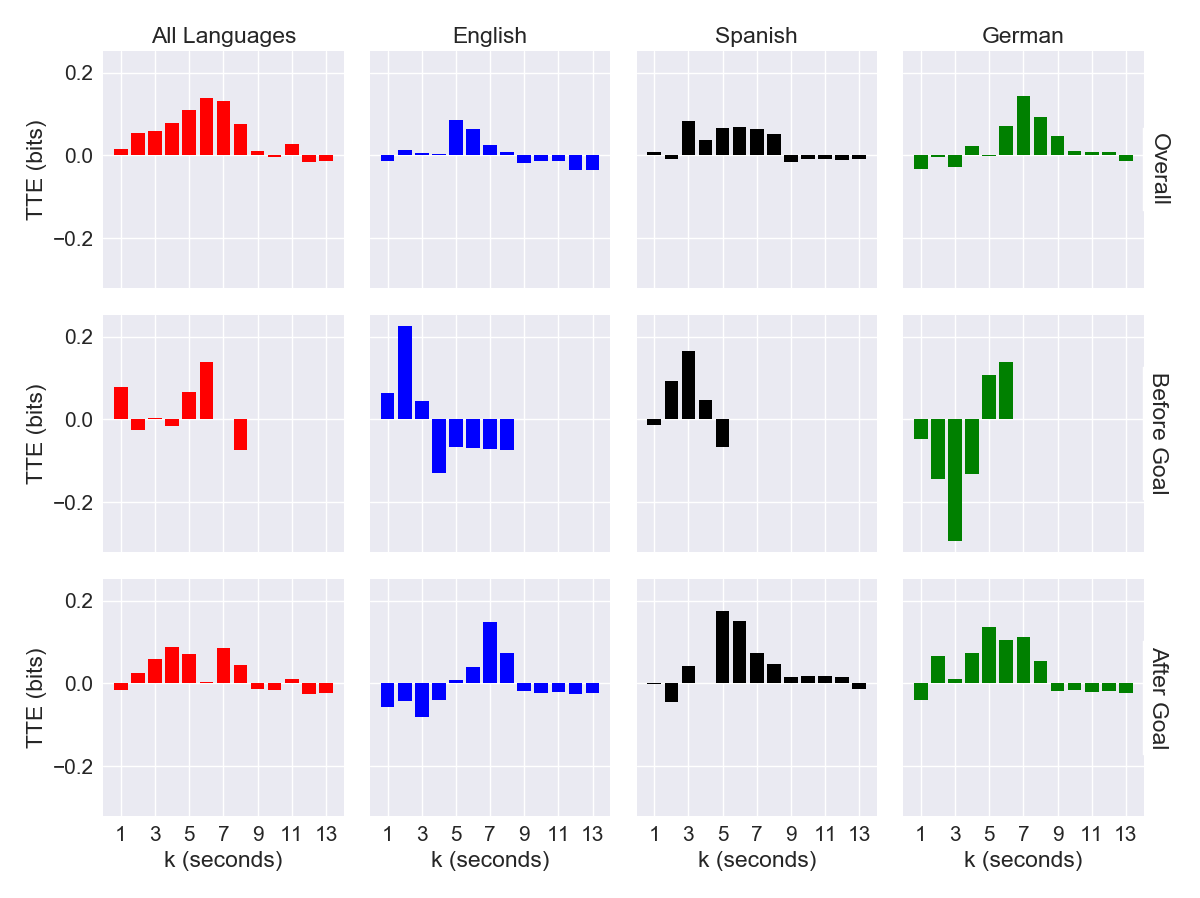}
\caption{$TTE$ vs $k$ categorized by language (columns), before and after the goal was scored by Mexico (rows). The goal clearly caused a shift in the value of $k$ for which maximum $TTE$ was observed for all three languages. There is an increased sentimental resilience shown by English and Spanish speakers and a reduced sentimental resilience shown by German speakers following the goal by Mexico. This is indicated by the increase in $k$ for maximal $TTE$ for English and Spanish speaking groups and a decrease for the German speaking group, respectively.}
\label{fig:TTE}
\end{figure}

Several relationships were found between the sentiment of the commentary and the endogenous sentiment of the Twitter discussions over time, which are summarized in Fig. \ref{fig:TTE} and discussed below.

\subsubsection{Entire population of people posting tweets about the game}

The TTE from the sentiment time-series of the transcript to the sentiment time-series of all tweets is shown in the first column (All Languages) of Fig. \ref{fig:TTE}. The overall TTE was maximum at $k=6$ with $TTE = 0.1386$ bits. Whereas, TTE is maximum at $k=6$ ($TTE = 0.1379$ bits) before and $k=4$ ($TTE = .0884$ bits) after the goal was scored, respectively. This provides evidence that rate of information flow from the exogenous events of the game into the online discussions was quickened after the goal was scored. 

\subsubsection{Tweets posted in different languages}
Column 2 to 4 of Fig. \ref{fig:TTE} present the TTE observed when the sentiment timeseries were categorized by the languages English, Spanish, and German. Overall, $TTE$ was maximum at $k=5$, $k=3$, and $k=7$ for English, Spanish, and German speaking fans, respectively. The maximum information flow from the exogenous events into each of these groups was $TTE=0.0851$, $TTE=0.0829$ and $TTE=0.1426$ bits, respectively.

Interestingly, we observe a clear shift in the rate of information flow after the goal is scored for each group. The $k$ for which English and Spanish speaking groups have maximum $TTE$ increases after the goal, from $k=2$ to $k=7$ for English and $k=3$ to $k=5$, indicating that the information flow from commentary sentiment to conversation sentiment was slowed. In contrast, $k$ for maximum $TTE$ dropped from $k=6$ to $k=5$, with lower values of $k$ generally giving higher $TTE$ values after the goal was scored. This indicated that the sentiment of German speaking fans was more sensitive to the sentiment of the commentary after the goal was scored. For German speaking fans, before the goal was scored, for $k<=4$, $TTE$ was generally less than 0. This could indicate that there was an increase in uncertainty when considering this history length. Further analysis is required to establish this.

Overall, assuming Spanish speaking fans supported the Mexico team and German speaking fans support the German team, these results indicate that following the exogenous shock, the sentiment of the group that was advantaged was more resilient and less susceptible to exogenous events, while the sentiment group that was disadvantaged was more sensitive and susceptible to exogenous events.

\section{Conclusion}
This paper explores a case where the discussions within and between a polarized population of two competing groups are affected by exogenous events. We consider the case of Twitter conversations and commentary on the Mexico vs Germany FIFA 2018 qualifier game, in which a single goal was scored by the Mexican team. This context presents us a unique situation where two polarized competing groups can be studied in isolation.
We investigate the effect that game events have on the conversation dynamics, namely, the volume, virality, and the responsiveness of the  conversations related to the game. Further, we investigate the emotional influence that the commentary had on the users tweeting about the game. 

We discover that while volume and virality were less influenced by the game events and more by the popularity of the root users, responsiveness to conversations was more influenced by the game events. 

Interestingly, we find that following the exogenous shock, the advantaged group showed increased emotional resilience to further exogenous events. In contrast, the disadvantaged group was left overall more emotionally susceptible to further exogenous events. These results imply that endogenously grown competitive discussions between two polarized groups that are emotionally invested in a particular topic may be used to manipulate and, possibly, alter the stability of one of the groups involved, while increasing the resilience of the other. If taken to its extreme, these results could serve as a warning to policy makers against the potential harm that could be caused to a population through targeted exogenous events focusing on a topic on which both groups are highly emotionally invested.

\begin{acknowledgement}
The authors would like to thank the Complex Systems Summer School 2018 conducted by the Santa Fe Insitute, at which this project was conducted and without which this collaboration would not have been possible.
\end{acknowledgement}

\bibliographystyle{apacite}
\bibliography{sample-base}

\end{document}